\documentclass[a4paper,11pt]{article}
\pdfoutput=1 

\usepackage{jheppub} 

\usepackage[T1]{fontenc} 
\usepackage{amsmath}
\usepackage{amsfonts}
\usepackage{amsmath}
\usepackage{amsxtra}
\usepackage{amstext}
\usepackage{amssymb}
\usepackage{slashed}
\usepackage{graphicx}
\usepackage{cleveref}
\usepackage{color}
\usepackage[utf8]{inputenc}

\numberwithin{equation}{section}

\newcommand{\beq}{\begin{equation}}
\newcommand{\eeq}{\end{equation}}
\newcommand{\bea}{\begin{eqnarray}}
\newcommand{\eea}{\end{eqnarray}}

\newcommand{\ep}{\epsilon}

\newcommand{\nn}{\nonumber}
\newcommand{\<}{\langle}
\renewcommand{\>}{\rangle}

\newcommand{\im}{\mathrm{Im}\,}
\newcommand{\re}{\mathrm{Re}\,}
\newcommand{\z}{\zeta}

\DeclareMathOperator{\Tr}{Tr}

\newcommand{\Hb}{\tilde H}
\newcommand{\lala}{\lambda \lambda}

\title{\boldmath Gaugino mass term for D-branes and Generalized Complex Geometry}

\author[a]{Mariana Gra\~na,}
\author[b]{Nicol\'as Kovensky,}
\author[a]{Ander Retolaza.}

\affiliation[a]{Institut de Physique Th\'eorique, Universit\'e Paris Saclay, CEA, CNRS, Orme des Merisiers, 91191 Gif-sur-Yvette Cedex, France.}
\affiliation[b]{Mathematical Sciences and STAG Research Centre, University of Southampton, Southampton, SO17 1BJ, United Kingdom.}

\emailAdd{mariana.grana@ipht.fr}
\emailAdd{n.kovensky@soton.ac.uk}
\emailAdd{ander.retolaza@ipht.fr}

\abstract{We compute the four-dimensional gaugino mass for a D$p$-brane extended in spacetime and wrapping a cycle on the internal geometry in a warped compactification with fluxes. Motivated by the backreaction of gaugino bilinear VEVs, we use Generalized Complex Geometry to characterize the internal geometry as well as the cycle wrapped by the brane.  We find that the RR fluxes and the non-closure of the generalized complex structures combine in the gaugino mass terms in the same form as they do in the bulk superpotential, while for the NSNS fluxes there is a crucial minus sign in the component normal to the brane. Our expression extends the known result for D3 and D7-branes in Calabi-Yau manifolds, where the gaugino masses are induced respectively by the imaginary anti-self dual and imaginary self-dual components of the complex 3-form flux $G_3$. 
}

\preprint{IPhT-t20/018}

\begin{document} 
\maketitle
\flushbottom

\section{Introduction and main result}
\label{sec:intro}

Gaugino bilinear vacuum expectation values (VEVs) play a prominent role in the mechanism of moduli stabilisation, both in heterotic and type II theories.
In the latter, where the gaugini are part of the fermionic degrees of freedom on D$p$-branes, such VEVs give rise to non-perturbative contributions to the effective potential involving the modulus that corresponds to the size of the cycle wrapped by the D-brane. In compactifications of type IIB theory, where D$p$-branes wrap cycles of even dimension, there are no perturbative contributions to the potential for these moduli coming from NSNS or RR fluxes
, and the non-perturbative terms become the leading order contribution. This is the mechanism proposed by Kachru, Kallosh, Linde and Trivedi (KKLT) \cite{Kachru:2003aw} to stabilise the K\"ahler moduli (corresponding to sizes of four-cycles) in type IIB compactifications on Calabi-Yau manifolds. 

The starting point in the KKLT construction is  the well-known set-up with a self-dual combination of NSNS and RR 3-form fluxes such that the background geometry is Calabi-Yau, with a metric that is related to the Ricci-flat metric by a conformal factor \cite{Grana:2000jj,Giddings:2001yu}. In the effective four-dimensional theory corresponding to the compactification on a Calabi-Yau manifold with 3-form fluxes, this solution appears as supersymmetric or supersymmetry-breaking Minkowski vacuum with stabilised complex structure moduli, which measure the sizes of 3-cycles. However, the K\"ahler moduli corresponding to sizes of 4-cycles are flat directions, which in this construction
are lifted by gaugino bilinear VEVs on D7-branes wrapping these cycles. The resulting effective field theory has supersymmetric  AdS$_4$ vacua with all moduli stabilised. Furthermore, these vacua exhibit a clear separation between the KK scale and the AdS curvature, allowing for a four-dimensional truncation of the field theory. Given the problems encountered in reproducing this feature in classically stabilised vacua (see e.g. \cite{Douglas:2006es,Tsimpis:2012tu,Petrini:2013ika,Gautason:2015tig,Gautason:2018gln,Font:2019uva}), this makes gaugino condensates interesting not only for KKLT but also for any type II compactification.

However, same as fluxes, gaugino bilinear VEVs back-react on the geometry: the internal manifold cannot be Calabi-Yau (or conformally related to it), since it does not support AdS$_4$ vacua. 
Moreover, it has been shown \cite{Koerber:2007xk,Dymarsky:2010mf,Heidenreich:2010ad,Bena:2019mte} that it cannot even be a geometry with SU(3) structure (defined by a single globally defined spinor), but it requires a more general ``dynamic SU(2) structure" (with two globally defined spinors that can become parallel at points or subspaces of the internal manifold). This situation is better described in the language of Generalized Complex Geometry (GCG,  see \cite{Koerber:2010bx} for a review), in terms of SU(3)$\times$SU(3) structures on the tangent plus cotangent bundle of the manifold. In order to try and describe backgrounds with gaugino bilinear vevs (i.e. a VEV for the operator corresponding to the gaugino mass terms) on D-branes, one is thus led to consider the generalized (almost) complex geometry of the internal manifold and that of the cycle wrapped by the brane.

In this paper we compute the four-dimensional gaugino mass term for a space-filling D$p$-brane, which wraps a cycle $\Sigma$ on the internal manifold. The brane is stable if the cycle satisfies the so-called calibration conditions, which split into algebraic and differential requirements. We assume that the cycle satisfies the former, but not necessarily the latter. For example, for D7-branes in SU(3)-structure backgrounds characterized by a real  (1,1)-form $J$ and a holomorphic (3,0)-form $\Omega$, this means that we assume the four-cycle to be holomorphic with respect to the almost complex structure of the background. On the other hand, we do not require the differential calibration conditions. In the D7-brane example, this implies we do not necessarily assume the complex structure to be integrable (or in other words the geometry need not be complex, that is, $d\Omega$  can be anything), nor do we assume any relation between $d\left(J^2\right)$ and the 5-form flux. 

To compute the gaugino mass term we use the fermionic D$p$-brane action at the two fermion level of \cite{Grisaru:2000ij,Marolf:2003vf,Marolf:2003ye,Martucci:2005rb}. This is the action for a single D$p$-brane\footnote{The non-Abelian version of the fermionic action is not known. Up to order $\alpha'^2$ the non-Abelian generalization of the fermionic DBI action amounts to adding a trace to the abelian expression \cite{Bergshoeff:2001dc}. At higher orders one expects though the appearance of new terms that are absent in the Abelian case (see \cite{Hamada:2018qef,Kallosh:2019oxv} for recent discussions on possible couplings at four-fermion level).} with world-volume flux embedded in a generic bosonic flux background. Doubling the fermionic degrees of freedom,  the quadratic fermionic action can be written compactly in a canonical Dirac-like form. The redundant degrees of freedom are removed by choosing a gauge for the fermionic  $\kappa$-symmetry of the action. For simplicity, we set the world-volume fluxes to zero.  
    
Let us state the main result of this paper. We find that, for a generic D$p$-brane in an SU(3)$\times$SU(3) structure background described in terms of the even and odd pure spinors (or polyforms) $\Psi_+$ and $\Psi_-$, the gaugino mass is given by  
\beq
m_{\lambda}=   -\dfrac{i}{8\pi}  \int_{M_6} \delta^{(0)}  \left\langle    \Psi_{\pm}  \, , \,   {F} + i d_{\Hb} (e^{-\phi} \re \Psi_{\mp})  \right\rangle  \nn \ ,
\eeq
where the upper (lower) sign is for type IIA (IIB). Here $\delta^{(0)}$ is a scalar delta-function with support at the locus of the brane\footnote{For example for a D7-brane in SU(3) structure it is the contraction of the 2-form 
 Poincare dual to the 4-cycle, and $J$. The general definition is presented below in \eqref{delta0general}.},
$\langle , \rangle$ denotes the Mukai pairing, defined in \eqref{Mukai}, $F$ is the even (odd) polyform given by the formal sum of internal RR fluxes in type IIA (IIB)   
and $ d_{\Hb}$ is the twisted exterior derivative where the two components of $H$ flux with none and 2 world-volume indices appear with a relative sign (the definition is written in \eqref{dHb}).

The calibration conditions imply that many of the components of the pure spinors vanish on the brane locus, such that only some of the components of the fluxes give a non-zero contribution to the gaugino mass. For instance, for a D3-brane, only the RR 3-form flux $F_3$ leads to a mass term, irrespectively of the (generalized) structure of the background.   

The contribution from the RR fluxes and the derivative of the pure spinors to the gaugino mass is of the same form as in the bulk superpotential \cite{Grana:2006hr,Benmachiche:2006df}, given in \eqref{W10D}. 
However, the NSNS flux enters with a slight (but key) difference: there is a relative sign in the component  along the normal directions to the brane compared to the twisted exterior derivative that appears in the superpotential, or in the supersymmetry equations, given in \eqref{dH}. This relative sign is such that for instance for D3-branes, where the derivative term gives no contribution, the gaugino mass is proportional to $\bar G_3 \wedge \Omega$, in accordance with \cite{Grana:2002tu,Camara:2003ku,Grana:2003ek}. For D7-branes instead, the gaugino mass involves the component of $H$ with two directions along the world-volume, which does not have a relative minus sign, and thus the gaugino mass is proportional to $G \wedge \Psi_-$, with $G$ the complex polyform defined in \eqref{Ggen}, and this reduces in Calabi-Yau manifolds to $G_3 \wedge \Omega$, in agreement with \cite{Jockers:2004yj,Camara:2004jj}. Thus imaginary self dual fluxes as in the solutions \cite{Dasgupta:1999ss,Grana:2000jj,Giddings:2001yu} do not generate gaugino masses on D3-branes, but they do on D7. The latter is also true in D8-branes, since there are no components of $H$ completely normal to the brane.
In D4-branes, there is only the contribution from the normal component to the brane, which has a minus sign, but since the derivative term is not zero, the NSNS contribution cannot be combined with the RR piece into a $\bar G$ as for D3-branes. For D5 and D6 branes both components of $H$-flux enter, and similarly we cannot write generically the mass term in terms of $G$ and/or $\bar G$.

The paper is organized as follows. In section \ref{sec:GG} we introduce the GCG description of type II compactifications, 
and also review how space-time-filling D-branes wrapping calibrated internal cycles are included in this framework. In section 
\ref{sec:WVaction} we use this formalism to compute the gaugino mass terms. We first introduce the quadratic fermionic D-brane 
action and set our conventions for the dimensional reduction. We then present the computation in some 
detail and give our final result in Eqs \eqref{mlambda}-\eqref{mfinalalternativo}. Finally, we show how the gaugino mass looks like for the particular 
case of an SU(3) structure geometry. Further conventions are presented in Appendix \ref{app:coventions} together with some useful identities.

\section{Compactifications, branes  and Generalized Complex Geometry}
\label{sec:GG}

\subsection{Compactifications and GCG}

We start with type II superstring theory on a warped product of an extended and maximally symmetric four-dimensional manifold (Mink$_4$, AdS$_4$ or dS$_4$) and a compact internal six-dimensional manifold $M_6$. The metric ansatz is %
\beq \label{10dmetric}
ds_{10}^2=e^{2A(y)} g^4_{\mu\nu}(x) dx^\mu dx^\nu + g^6_{mn}(y) dy^m dy^n \ ,
\eeq
where $x^\mu, \mu=0,..3$ are external coordinates while $y^m$, $m=1,..6$ are coordinates on $M_6$. In order to preserve isotropy and homogeneity in the extended directions the NSNS field-strength $H$ can only have all internal legs. In the democratic formulation, this forces the RR fluxes to be of the form
\beq
\label{Fintext}
F = {\hat F} + e^{4A} \mathrm{vol}_4 \wedge \tilde{F} \ .
\eeq
Here we use the democratic formulation of \cite{Bergshoeff:2001pv} (mostly following the conventions of \cite{Lust:2008zd}) and the polyform notation such that $F = \sum F_q$ with $q=0,2,4,6,8,10$ ($q=1,3,5,7,9$) for type IIA (IIB).
Both $\hat F$ and $\tilde{F}$ have only internal legs, and the self-duality condition for $F$ can be brought to the more useful form
\beq \label{sdF6}
\tilde F= *_6 \, \alpha({\hat F}) \ ,  \quad  \alpha(\omega_q)=(-1)^{\frac{q(q-1)}{2}}\omega_q.
\eeq
Note that the operator $*_6 \circ \alpha$ squares to $-1$. 
We require the background to have globally defined spinor(s)\footnote{This condition is necessary for the theory on the branes to be a (softly broken) ${\cal N}=1$ gauge theory.}. In type II string theory, the most generic situation is to consider two globally defined spinors 
$\eta^1$ and $\eta^2$, which can become parallel at certain loci of the manifold. 
By using these spinors, one can build two polyforms or pure spinors which characterise the background geometry, and are defined as 
\begin{eqnarray}
\label{purespinors}
    \Psi_{+} = -\frac{ i}{||\eta||^2} \sum_{p \text{ even}} \frac{1}{p!} \eta^{2\dagger}_{+} \gamma_{m_1 \dots m_p} \eta^1_+ \, dy^{m_1} \wedge \dots \wedge dy^{m_p} \ , \quad 
    \underline{\Psi_+} = 
    -\frac{8 i}{||\eta||^2}
    \eta^1_+ \otimes \eta^{2 \, \dagger}_+  , \nn\\
 \Psi_{-} = -\frac{ i}{||\eta||^2} \sum_{p \text{ odd}} \frac{1}{p!} \eta^{2\dagger}_{-} \gamma_{m_1 \dots m_p} \eta^1_+ \, dy^{m_1} \wedge \dots \wedge dy^{m_p} \ , \quad 
    \underline{\Psi_-} = 
    -\frac{8 i}{||\eta||^2}
    \eta^1_+ \otimes \eta^{2 \, \dagger}_- ,
\end{eqnarray}
where the subindex on the interal spinors $\eta^i$ denotes their chirality, and 
 underlined forms are contracted with $\gamma$-matrices as defined in   (\ref{eq:def-underline}). These bi-spinors define an SU(3)$\times$SU(3) $\subset \mathrm{Spin}(6,6)$ structure.

In the well-known case where the internal manifold accepts only one globally well-defined spinor   $\eta^1$ and $\eta^2$ are parallel everywhere. This is  known as an SU(3) $\subset$ O(6) structure compactification. In such configurations the pure spinors reduce to 
\begin{equation} \label{SU3ps}
    \Psi_-= i e^{- i \theta} \Omega \ , \quad  \Psi_+=    e^{i \theta} \exp (i J) \ .
\end{equation}
Here $J$ and $\Omega$ are the usual real (1,1)-form and holomorphic (3,0) form defining the SU(3) structure, and  $\theta$ is the relative phase between the spinors: $\eta^1_+= ie^{ i\theta} \eta^2_+$. In our conventions, described in detail in Appendix \ref{app:coventions}, a supersymmetric background compatible with D3/D7-branes has $\theta=0$. Similarly, for the D5/D9 supersymmetry we take  $\theta=-\pi/2$. 

\medskip

It is not hard to show that the pure spinors satisfy the self-duality condition
\beq \label{*Psi}
*_6 \alpha(\Psi) = i \Psi \ . 
\eeq
We also introduce for later use the Mukai pairing, which is the form version of the inner product between Spin(6,6) spinors 
\begin{equation} \label{Mukai}
    \<\Psi,\Phi\> \equiv \left[\Psi \wedge \alpha(\Phi) \right]_{6} \ ,
\end{equation}
where $[\cdot ]_{6}$ indicates one should only keep the 6-form in the polyform wedged product. Moreover, the pure spinors $\Psi_+$ and $\Psi_-$ are ``compatible'', which means that the following identities hold:
\beq \label{compat}
\langle \Psi_+ , dy^m \wedge \Psi_- \rangle = \langle \Psi_+ , \iota_m  \Psi_- \rangle=0 \ , \
\langle \Psi_+, \bar \Psi_+ \rangle = \langle \Psi_-, \bar \Psi_- \rangle = -8 i \, {\rm vol}_6. 
\eeq

\paragraph{Supersymmetric backgrounds:}
The background preserves ${\cal N}=1$ supersymmetry if the supersymmetry variations of the gravitino $\psi$ and the dilatino $\lambda$\footnote{We work in the same fermionic frame as \cite{Bergshoeff:2001pv}, where there is a crossed kinetic term mixing dilatino and gravitino. Our results are nevertheless independent of the frame chosen since we are interested in bosonic backgrounds. }
\begin{eqnarray} \label{eq:operator1}
\delta_\epsilon \psi_M=D_M\epsilon=\left(\nabla_M +\dfrac{1}{4} \underline{H_{M}}\sigma_3  +\dfrac{e^{\phi}}{16}  \sum_q \underline{F_q} \Gamma_M \mathcal{P}_q \right)\epsilon \ , \\
\delta_\epsilon\lambda=\Delta\epsilon=\left(\underline{d\phi} +\dfrac{1}{2} \underline{H}\sigma_3  +\dfrac{e^{\phi}}{16}  \sum_q \Gamma^M\underline{F_q} \Gamma_M \mathcal{P}_q \right)\epsilon \  \label{eq:operator2}
\end{eqnarray}
vanish for a given supersymmetry parameter $\epsilon$. We use the double spinor notation, and $\sigma_3$ as well as $\mathcal{P}_q=\left\lbrace  - \sigma_1, i \sigma_2 , -i \sigma_2 ,\sigma_1 \right\rbrace$ for $q=\left\lbrace  0,1,2,3\right\rbrace$ mod $4$ act on the fermion doublets. Finally, 
$\underline{H_{M}}$ is defined in \eqref{HM}.

Since the internal manifold has two globally defined spinors, it is natural to use them in the decomposition of the supersymmetry parameter into an external and an internal spinor, namely  
\begin{equation}
 \label{epsilon12}
 \ep\equiv  \left(\begin{matrix}
\ep^1_+ \\ \ep^2_\mp
\end{matrix}\right)= \zeta_+ \otimes \left(\begin{matrix}
\eta^1_+ \\ \eta^2_\mp
\end{matrix}\right)   + \zeta_- \otimes 
\left(\begin{matrix}
\eta^1_-\\ \eta^2_\pm
\end{matrix}\right) \ 
\end{equation}
 (see conventions for the fermions in Appendix \ref{sec:spinorconventions}) where $\epsilon^{1}_+$ is a 10d Majorana-Weyl spinor of positive chirality, and $\epsilon^2_\mp$ has the opposite (same) chirality in type IIA (IIB) and  $\z_+$ and $\z_-$ are 4d Dirac spinors
required to satisfy $2 \nabla_\nu \z_+ = \pm \bar{\mu} \gamma_\mu \z_-$, with $\mu$ related to the 4d cosmological constant by $\Lambda = -3|\mu|^2$. Here and in the rest of the paper the upper (lower) sign corresponds to type IIA (IIB).

The supersymmetry conditions are equivalent to the following differential equations on the pure spinors \cite{Grana:2004bg,Grana:2005sn}
\begin{subequations}
\label{susy}
 \begin{align}
d_H \left(e^{3A-\phi} \Psi_2\right) &= 2 i \mu e^{2A-\phi} \im \Psi_1  \label{susy1}\\
d_H \left(e^{2A-\phi} \im \Psi_1 \right) &= 0 \label{susy2}\\
d_H \left(e^{4A-\phi} \re \Psi_1 \right) &= 3 e^{3A-\phi} \re \left[ \bar{\mu} \, \Psi_2 \right] + e^{4A} *_6 \alpha(\hat F) \label{susy3}
 \end{align}
\end{subequations}
where 
\beq \label{dH}
d_H = d + H\wedge \ 
\eeq
is the twisted exterior derivative and
\beq
\Psi_1=\Psi_\mp \ , \quad \Psi_2=\Psi_\pm \ . 
\eeq

Importantly, it was shown in \cite{Grana:2005ny,Grana:2006hr,Benmachiche:2006df,Koerber:2007xk,Koerber:2008sx}  that the above supersymmetry conditions can be obtained as D- and F-flatness conditions from the superpotential 
\begin{equation}
    W =  
 \pi   \int_{M_6} \<e^{3A-\phi} \Psi_2,  d_H [\hat{C} + i e^{-\phi} \mathrm{Re} \Psi_1]\> 
    = \pi \int_{M_6} \< e^{3A-\phi} \Psi_2 , G\> \ .
    \label{W10D}
\end{equation}
Here $\hat{C}$ are the internal RR gauge potentials ($d_H \hat{C} = \hat F $), and $G$ 
is the polyform that generalizes the $G_3$-flux \cite{Lust:2008zd}
\begin{equation}   \label{Ggen}
 G \equiv \hat F + i e^{-4A} d_H \left(e^{4A - \phi} \re \Psi_1\right) \ .
\end{equation}
It follows from \eqref{susy3} that for supersymmetric GKP compactifications, $G = G_3$ satisfies the usual ISD condition, while in more general supersymmetric compactifications this generalizes to 
\begin{equation}  \label{genISD}
     (1- i *_6 \alpha  ) \, G =  
    3 i  e^{-A-\phi}\bar{ \mu} \Psi_2\  .
\end{equation}
 
For $\mu=0$, the ISD requirement \eqref{genISD} also describes a class of supersymmetry breaking solutions with $(0,3)$ three-form flux $G_3$ \cite{Grana:2000jj}, or more generally a component of $G$ along $\bar \Psi_2$ \cite{Lust:2008zd}. 

\subsection{D-branes in GCG}

Consider a space-time filling D$p$-brane, wrapping a cycle $\Sigma_{p'}$ ($p'=p-3$) on the internal manifold whose geometry is encoded in the pure spinors. As is well known, for the D-brane to be stable, $\Sigma_{p'}$ has to be a minimal volume hyper-surface, denoted as calibrated submanifolds for BPS objects. In this situation, the supersymmetry generators\footnote{We call $\epsilon$ the supersymmetry generators even if the background is not necessarily supersymmetric, i.e. we do not require \eqref{eq:operator1} or \eqref{eq:operator2} to vanish.} satisfy at the brane locus
\beq \label{GammaDpepsilon}
\left[ \bar{\epsilon}_1 =- \bar{\epsilon}_2 \Gamma_{D_p} \right]_{\Sigma}
\eeq 
where $\Gamma_{D_p}$ 
 is defined in \eqref{GammaDp}. We review this setup in the context of generalized complex geometry, following \cite{Koerber:2010bx}. 
 
To be precise, in the GCG context we should in fact talk about generalized submanifolds and calibrations \cite{Koerber:2005qi}, whose definitions take into account the background $B$-field and the gauge field  $A_1$  on the brane. The world-volume fluxes are characterized by the gauge invariant combination\footnote{In this expression one should write $P[B_2]$, where $P[]$ stands for the pullback of a background field to the internal cycle $\Sigma_{p'}$. Here and in what follows we simply leave the pullback operation for background fields implicit, unless otherwise stated. } 
 ${\cal{F}} \equiv B_2 + (2 \pi)^{-1} dA_1$ 
such that the generalized submanifold is given by the pair ($\Sigma_{p'},{\cal{F}}$). 
The bosonic part of the D$p$-brane action reads 
\begin{equation} 
\label{SDp}
\mathcal{S}^{B}_{D_p} = -2 \pi  \int d^{p+1} \xi e^{-\phi} \sqrt{ -\text{det} (g +\mathcal{F})}  +2 \pi\int C \wedge e^\mathcal{F},
\end{equation}
such that for a space-time filling brane the integration is over the warped product of $X_4$ and $\Sigma_{p'}$, and we set $2\pi\sqrt{\alpha'}=1$, such that $T_{p}= 2\pi $  is the D$p$-brane tension and $C$ is the sum of the RR potentials, i.e. $d_H C = F$. 

The generalization of the minimal surface condition then tells us that the ($\Sigma_{p'},{\cal{F}}$) is calibrated iff the DBI energy density (per unit of the  generalized volume) can be written as 
\begin{equation}
    d^{\,p'} \xi \, e^{4A- \phi }\sqrt{\text{det} (g_6 + \mathcal{F})} = [e^{4A - \phi} \re \Psi_1\rvert_{\Sigma} \wedge e^{\mathcal{F}}]_{p'},
    \label{calibration1}
\end{equation}
and   the following differential condition holds: 
\begin{equation}
\label{calibration2}
d_H \left( e^{4A - \phi} \re \Psi_1 \right)=3 e^{3A-\phi} \re \left[ \bar{\mu} \, \Psi_2 \right] + e^{4A} \tilde F \ . 
\end{equation}
This can also be obtained by studying the corresponding supersymmetry conditions. In fact, $e^{4A - \phi} \re \Psi_1$ constitutes a so-called generalized calibration form, and \eqref{calibration1} is the algebraic calibration condition that we require. 

In this paper we require the brane to satisfy the algebraic calibration condition \eqref{calibration1}, but not the differential one \eqref{calibration2}. Note that the latter coincides with the bulk supersymmetry equation \eqref{susy3}. 
 The remaining supersymmetry equations, \eqref{susy1} and \eqref{susy2}, which generically we do not impose either,  can also be interpreted as calibration conditions for domain wall-type and string-like D-branes respectively, as seen from the four-dimensional perspective.

Some well-known examples of calibrated submanifolds can easily be described for compactifications with internal SU(3) structure  in the $\mathcal{F}=0$ case. 
On the one hand, there are the $2l$-dimensional complex submanifolds, related to the calibration forms proportional to $J^{l}\sim [\re \Psi_1]_{2l}=\re [(e^{i\theta} e^{iJ})]_{2l}$ in type IIB, where $\theta$ depends on the brane in question as explained below \eqref{SU3ps}. On the other hand, there are the special Lagrangian submanifolds. In this case the calibration form is $\re (e^{-i\theta} \Omega)$. 

\medskip

It will prove useful to describe the calibration condition \eqref{calibration1} in more detail. It can be split as follows \cite{Martucci:2005ht} (see also \cite{Martucci:2006ij}). First, one has
\begin{equation}\label{eq:impsi1}
    \left[\im  \Psi_1    \wedge e^{\mathcal{F}}\right]_{\Sigma_{p'}}
    =0.
\end{equation}
This is nothing but the D-flatness condition for the four-dimensional effective theory on the D-brane. Second, one requires that ($\Sigma_{p'},{\cal{F}}$) is a generalized complex submanifold with respect to the generalized complex structure associated to the other pure spinor, $\Psi_2$. This corresponds to the 4d F-flatness condition, and implies that 
\begin{equation}\label{eq:calibrationspinors}
         \left[(dy^m \wedge \Psi_2 + g^{mn}\imath_n \Psi_2)\wedge e^{\mathcal{F}}\right]_{\Sigma_{p'}} = 0 \ . 
\end{equation}
 For instance, in the special Lagrangian example given above, the D- and F-flatness conditions are actually the \textit{special} and \textit{Lagrangian} conditions, respectively. 
 These calibration conditions will be crucial for our computation, so we will provide further details about them shortly.

Finally, we define $\delta^{6-p'}[\Sigma_{p'}]$ as the $(6-p')$-form Poincare-dual to the cycle $\Sigma_{p'}$ where a D$p$-brane is wrapped, namely 
\begin{equation}
 \int_{\Sigma_{p'}} \omega  \equiv  \int_{M_6} \omega \wedge \delta^{6-p'}[\Sigma_{p'}]  
\end{equation}
for any $p'$-form $\omega$. We also define $\delta^{0}[\Sigma_{p'}]$ as the associated scalar $\delta$-function in the form
\beq \label{delta0general}
\delta^{0}[\Sigma_{p'}]\equiv (*{\rm vol}_{\Sigma_{p'}})^{6-p'} \cdot  \delta^{6-p'}[\Sigma_{p'}] \quad \Rightarrow \quad \delta^{0}[\Sigma_{p'}] \, {\rm vol}_6 = \< {\rm vol}_{\Sigma_{p'}}, \delta^{6-p'}[\Sigma_{p'}]  \> \ ,
\eeq
here ${\rm vol}_{\Sigma_{}p'}$ is the volume form on the cycle and we used the form scalar product defined in \eqref{eq:scalarproduct}. If the brane satisfies the algebraic calibration condition \eqref{calibration1}, this is equivalent to 
\beq \label{delta0calibrated}
\delta^{0}[\Sigma]= (\im \Psi_1)^{6-p'} \cdot \delta^{6-p'}[\Sigma] \quad \Rightarrow \quad \delta^{0}[\Sigma] \, {\rm vol}_6 = \< \re \Psi_1, \delta^{6-p'}[\Sigma]  \>
\eeq
where $(\im \Psi_1)^{6-p'}$ is the $6-p'$ vector dual to the $6-p'$ form-piece of $\im \Psi_1$.

\section{Fermionic D-brane action}
\label{sec:WVaction}

The supersymmetric version of the DBI and Wess-Zumino action at the two-fermion level was computed in \cite{Marolf:2003ye,Marolf:2003vf,Martucci:2005rb}. For any D$p$-brane, the bosonic terms were given in \eqref{SDp} and  the fermionic terms can be compactly written as
 \begin{equation}\label{SF}
\mathcal{S}^{F}_{D_p} =   i \pi \int d^{p+1} \xi e^{-\phi} \sqrt{ -\text{det} (g+\mathcal{F})} \bar{\theta } (1-\Gamma_\kappa   ( \mathcal{F})) \left( \Gamma^\alpha D_\alpha-\dfrac{1}{2}\Delta-L_{Dp} (\mathcal{F})\right)\theta \ . 
\end{equation}
  Here, the world-volume fermion  $\theta$ is a doublet of Majorana-Weyl (MW) fermions 
  \begin{equation}
\theta\equiv\left(\begin{matrix}
\theta_1\\ \theta_2
\end{matrix}\right) \quad , \quad \Gamma_{(11)}\theta = \left(\begin{matrix}
\theta_1\\ \mp \theta_2
\end{matrix}\right) \ ,  
\end{equation}
with $\Gamma_{(11)}$ the ten-dimensional chirality matrix. 
The $\theta_i$ are not independent: the fermionic action possesses a fermionic gauge symmetry called $\kappa$-symmetry. One can use this to gauge away half of the fermionic degrees of freedom. We fix the gauge by requiring
\beq \label{gf}
 \bar{\theta}\Gamma_\kappa   =-\bar{\theta} ,
 \eeq  
where $\Gamma_{\kappa}$ is built out of the world-volume chirality operator as in \eqref{eq:LF2}. This is in contrast with the supersymmetry generators defined in \eqref{epsilon12}, which satisfy \eqref{GammaDpepsilon} (in terms of $\Gamma_{\kappa}$, this equation reads $\bar{\epsilon}\Gamma_\kappa = \bar{\epsilon}$). The  sign difference with respect to \eqref{gf} is crucial since the WV spinor should not be proportional to the pullback of the supersymmetry generator, since for supersymmetric backgrounds the latter is a redundancy of the background. The matrix $\Gamma_\kappa$ squares to the identity, and thus $\tfrac12 (1-\Gamma_{\kappa})$ is a projector onto the subspace of ``physical" fermions.  The gauge-fixing condition \eqref{gf} relates the two MW fermions according to
\begin{equation} \label{kappagauge}
  \bar{\theta}_1=\bar{\theta}_2\Gamma_{Dp}\quad\quad \Leftrightarrow \quad\quad   \Gamma_{Dp } \theta_1=(-1)^{p} \theta_2 =\pm\theta_2 \ . 
\end{equation}
The operator $\Gamma_{Dp}$ is a `chirality' operator on the generalized world-volume of the D$p$-brane:\footnote{Note that our definition of $\Gamma_{Dp}$ has an extra minus sign as compared to the one in \cite{Marolf:2003vf,Marolf:2003ye,Martucci:2005rb}. 
} 
\begin{equation} \label{GammaDp}
\Gamma_{Dp} =\dfrac{(-1)^{[p/2]}}{\sqrt{-\det(g+\mathcal{F})}}\dfrac{\varepsilon^{\alpha_1...\alpha_{p+1}}\Gamma_{\alpha_1}...\Gamma_{\alpha_{p+1}}}{(p+1)!}\sum_{r=0}^{[p/2]}\dfrac{\Gamma^{\beta_1... \beta_{2r}}\mathcal{F}_{\beta_1\beta_2}... \mathcal{F}_{\beta_{2r-1}\beta_{2r}}}{2^r r!} \ ,
\end{equation}
here the indices run over world-volume directions and the matrices $\Gamma_{\alpha_i}$ represent the pullback of the $\Gamma_M$.
 The operator $L(\mathcal{F})$ is given in \eqref{eq:LF}, and vanishes for ${\cal F}=0$, which is the situation we will restrict to in this paper. 
Finally, $D_\alpha $ and $\Delta $ are the operators involved in the gravitino and dilatino supersymmetry variations, defined in  \eqref{eq:operator1} and \eqref{eq:operator2}. We are interested in the combination that appears in \eqref{SF}:
\begin{eqnarray}\label{eq:operators}
\hat D &\equiv& \Gamma^\alpha D_\alpha-\dfrac{\Delta}{2}  \\
&=&\Gamma^\alpha\nabla_\alpha-\dfrac{1}{2}\underline{d\phi}   + \dfrac{e^\phi}{16	} \sum_q \left[ \Gamma^\alpha \underline{F_q}\Gamma_\alpha - \frac{1}{2} \Gamma^M \underline{F_q} \Gamma_M \right]\mathcal{P}_q + \frac14 \left[\Gamma^\alpha\underline{H_{\alpha}}-\underline{H}  \right]  \sigma_{3} \ ,  \hspace{20pt}  \nn 
\end{eqnarray}
where  $q $ sums over all even RR field strengths, and ${\cal P}_q$ is defined above \eqref{epsilon12}.
 
 \subsection{Dimensional reduction: general idea}
  
In this section we consider a D$p$-brane in the warped compactifications described in section \ref{sec:GG}.
 The metric ansatz is given in \eqref{10dmetric}, bulk and brane fields split into external and internal components. The ten-dimensional $\Gamma$-matrices are decomposed as follows:    
\begin{equation} \label{gamma10}
 \Gamma^\mu =e^{-A} \gamma^\mu\otimes 1 \quad , \quad \Gamma^m= \gamma_{(5)}\otimes \gamma^m \quad , \quad \Gamma_{(11)}= \gamma_{(5)}\otimes \gamma_{(7)} \ , 
 \end{equation}
  $\gamma^\mu$ and $\gamma^m$ being the 4d and 6d gamma matrices while $\gamma_{(5)}$ and $\gamma_{(7)}$ represent the corresponding chirality operators. 
Let us now focus on the world-volume spinor $\theta$, which combines the degrees of freedom corresponding to the gaugino and the chiral fermions on the brane. Here we are only interested in the gaugino. Since we are considering D-branes that satisfy the algebraic calibration condition, we can use the bulk spinors $\eta^1$ and $\eta^2$ to write the 4d gaugino as \cite{Lust:2008zd}
\begin{equation} \label{thetalambda}
    \theta = 
    \left(\begin{array}{c}
      \theta^1 \\
      \theta^2 
    \end{array} \right) =  \frac{e^{-2A}}{4\pi} 
    \left( \begin{array}{c}
     \lambda_+\otimes \eta_+^1+ \lambda_-\otimes \eta_-^1    \\
     -\lambda_+\otimes \eta_\mp^2 - \lambda_-\otimes \eta_\pm^2
    \end{array} \right) + \cdots 
    \end{equation}
here  $\lambda_+$ and $\lambda_-$ are 4d Dirac spinors of definite chirality and 
 $+ \ ... \ $ stands for the 4d fermions in the three chiral multiplets. The overall numerical constant and warp factor are introduced in order to get  canonically normalized kinetic term for the 4d gauginos. 
The $\kappa $-fixing condition \eqref{gf} then implies 
\beq \label{eq:kappaintern}
 \left[ \eta^1_+  = i \gamma_{Dp'}\eta^2_\mp\right]_{\Sigma}  \ , \ 
 \gamma_{Dp'}=\dfrac{1}{\sqrt{\det g_6}}\dfrac{1}{p!} \, \varepsilon^{\alpha_1...\alpha_{p'}} \gamma_{\alpha_1}...\gamma_{\alpha_{p'}} \ ,
 \eeq
where the $\gamma$-matrices are internal. Note that the condition on the internal spinors, which is required to hold only at the brane locus, is the same as the internal part of the calibration condition \eqref{GammaDpepsilon}. The relative minus sign between  \eqref{GammaDpepsilon} and \eqref{kappagauge} has been absorbed when defining the 4d gaugino in \eqref{thetalambda}.

\subsubsection*{Calibration conditions revisited}

The relation \eqref{eq:kappaintern} between the internal spinors arises from the calibration condition \eqref{GammaDpepsilon} and can be used to better understand the local constraints \eqref{eq:impsi1}-\eqref{eq:calibrationspinors} that the pure spinors satisfy on calibrated cycles. 
For this, we first consider the $q$-form component of $\Psi_2=\Psi_\pm$ with $n$ indices along  the $p'$ cycle wrapped by a calibrated D$p$-brane. The calibration condition implies that 
\begin{equation}\label{eq:kappa-for-RR}
\left[ \eta^{2\dagger}_\pm \gamma_{m_1 ... m_q} \eta^1_+ = (-1)^{\frac{(p+2)(p+1)}{2}-n}\eta^{1\dagger}_- \gamma_{m_1 ... m_q} \eta^2_\mp= (-1)^{\frac{p+q}{2}+1-n}\eta^{2\dagger}_\pm \gamma_{m_1 ... m_q} \eta^1_+ \right]_\Sigma
\end{equation} 
where in the first equality we used \eqref{eq:kappaintern} twice  and in  the second one we took the transverse.  This implies that certain components of $\Psi_2$ are zero due to the calibration condition. An analogous calculation for $\Psi_1=\Psi_\mp$ shows that 
\begin{equation}\label{eq:another-kappa}
\left[ \eta^{2\dagger}_\mp \gamma_{m_1 ... m_q} \eta^1_+ = (-1)^{\frac{p(p+1)}{2}+1+q-n}\eta^{1\dagger}_+ \gamma_{m_1 ... m_q} \eta^2_\mp= (-1)^{\frac{p+q}{2}-\frac{1}{2}-n}(\eta^{2\dagger}_\mp \gamma_{m_1 ... m_q} \eta^1_+)^* \right]_\Sigma \ .
\end{equation} 
Therefore \eqref{eq:kappaintern} mixes $\Psi_1$ and its complex conjugate, so the calibration condition must be imposed separately on the real and imaginary parts of $\Psi_1$.

Note that in both cases the projection imposed by the calibration condition depends on the combination $p+q-2n$, which actually takes within certain range. Indeed, since $n$ indicates the amount of world-volume indices of the $q$-form component of a polyform, it  must satisfy
$0\leq n \leq q$. Also, because the brane wraps a $(p-3)$-cycle, necessarily $n\leq p-3$. Finally, requiring that there are enough transverse directions implies 
$q-n\leq 9-p$.
Combining these inequalities one finds that physically sensible options satisfy  $3\leq p+q-2n \leq 9$. This range can be combined with the above constraints to find that, {\it at the brane locus}, 
 \bea \label{eq:seispurespinors}
\Psi_2|_q^{(n)}&=& 0 \quad {\rm for }  \ p+q-2n\neq 6    \nn \\
\im \Psi_1|_q^{(n)}&=& 0 \quad  {\rm for } \  p+q-2n\neq 5 \ {\rm or} \ 9  \\ 
\re \Psi_1|_q^{(n)}&=& 0 \quad {\rm for } \  p+q-2n\neq 3 \ {\rm or} \ 7 \ .\nn
 \eea
 Note that the second and third line are equivalent due to the  self-duality property of the pure spinors \eqref{*Psi}. The conditions \eqref{eq:seispurespinors} are equivalent to \eqref{eq:impsi1}-\eqref{eq:calibrationspinors}, but they are written in a way that  will be useful later.

\subsection{Dimensional reduction: details of calculations}

Our goal is to compute the dimensional reduction of the fermionic action in order to obtain the gaugino mass-terms. We start with the gauge-fixed action  
\begin{equation}\label{SF46}
\mathcal{S}^{F}_{D_p} =  2  \pi i \int d^{p+1} \xi e^{-\phi} \sqrt{ -\text{det} g} \,  \bar{\theta }  \hat D \theta =  2  \pi i  \int_{X_4} d^4x  \sqrt{ -\text{det} g_4} \int_{\Sigma} e^{4A-\phi} \, \re \Psi_1 \bar{\theta }  \hat D \theta  \ 
\end{equation}
where $\theta$ satisfies \eqref{gf} and $\hat D$, defined in \eqref{eq:operators}, involves a derivative part, a term proportional to the gradient of the dilaton, a RR contribution and an NSNS flux piece. 
The derivative term involves both external and internal directions. The former gives the 4d gaugino kinetic term, while the latter combines with the fluxes and the dilaton giving the gaugino mass. The resulting four-dimensional action is then   of the form 
\beq \label{4daction}
{\cal S}_{\lambda\lambda}=\int d^4x \, \left( \frac{i}{2}f \, \bar{\lambda}_+\gamma^\mu\nabla_\mu\lambda_+ + \frac{1}{2} m_{\lambda} \,  
\bar{\lambda}_-\lambda_+ + {\rm c.c.} \  \right)  
\eeq
with $m_{\lambda}=m^d_{\lambda}+m_{\lambda}^\phi+m_{\lambda}^F +m_{\lambda}^H$
and $f$ the gauge kinetic function. 
 
 
Here we provide the detailed calculation of the dimensional reduction of the fermionic brane action. The uninterested reader may skip this section to  find the final result in the next one. We compute the kinetic function $f$, as well as each of the contributions to the mass separately.

\paragraph{Derivative term:} We need to compute 
\begin{equation}
   \bar{\theta} \Gamma^\alpha\nabla_\alpha \theta = \bar{\theta}_1\Gamma^\alpha\nabla_\alpha \theta _1+\bar{\theta}_2\Gamma^\alpha\nabla_\alpha \theta _2 
   \end{equation}
where $\alpha$ runs over   all world-volume directions. Separating external and internal indices we find\footnote{In \eqref{4daction} and \eqref{der1} the derivative $\nabla_\mu$  
denotes the covariant derivative with respect to the unwarped metric $g_{\mu\nu}$.} %
 \begin{eqnarray}    \label{der1}
     \bar{\theta}\Gamma^\alpha\nabla_\alpha \theta &=&\frac{e^{-5A}}{16\pi^2} \left[\bar{\lambda}_+\gamma^\mu\nabla_\mu\lambda_+ \otimes   (\eta^{1\dagger}_+\eta^1_+ + \eta^{2\dagger}_+\eta^2_+)  \ + \right.  \\
&& \quad   \quad \quad \qquad   \left.     \bar{\lambda}_- \lambda_+\otimes \left(\eta^{1\dagger}_-  \gamma^a\nabla_a  \eta^1_+ +\eta^{2\dagger}_\pm  \gamma^a\nabla_a  \eta^2_\mp \right)\right] + \, \mathrm{c.c.} \nn 
    \end{eqnarray} 
with $a=1, \dots, p'$. In the second line we have used the fact that the terms proportional to $dA$ are zero by virtue of \eqref{onegamma}. 
Taking the norm of the internal spinors to be $\eta^{1\dagger}_+\eta^1_+=\eta^{2\dagger}_+\eta^2_+=|\eta |^2=e^A$, as in supersymmetric compactifications, the first line gives the kinetic term in \eqref{4daction} with
\begin{equation}
\label{fkin}
    f = \frac{1}{2\pi}\int_\Sigma e^{-\phi} \text{ Re }\Psi_1 \ .
\end{equation}     
On the other hand, the derivative along the internal directions is slightly more involved. First, we note that using the $\kappa$-gauge fixing condition \eqref{eq:kappaintern} we get (no sum over $m$)
\begin{equation}\label{eq:kappaH}
   \eta^{1\,\dagger}_- \gamma^m \nabla_m \eta^1_+ 
   = (-1)^s \eta^{2\,\dagger}_\pm \gamma^m 
   \nabla_m \eta^2_\mp
\end{equation}
with $s=0$ if $m$ is a world-volume index and $s=1$ if it is transverse.
 This means that in the second line of \eqref{der1} we can actually sum over all internal indices. We then have 
\begin{eqnarray} \label{eq:blablabla}
    \eta^{1\dagger}_-  \gamma^m \nabla_m  \eta^1_+ +  \eta^{2\dagger}_\pm  \gamma^m \nabla_m  \eta^2_\mp &=&  e^{-A}\left[ \eta^{1\dagger}_-  \gamma^m (\nabla_m  \eta^1_+ \eta^{2\dagger}_\mp)\eta^{2}_\mp -\eta^{1\dagger}_- (\nabla_m \eta^{1}_-  \eta^{2\dagger}_\pm ) \gamma^m   \eta^2_\mp   \right]   \nn  \\  
    &=& \mp \frac{1}{2} \left(
    \Psi_2 \cdot d\, \re \Psi_1
    \right)
   \end{eqnarray}
where in the first line we have integrated by parts, and in the second one we used the ciclicity of the trace, the definition of the pure spinors   \eqref{purespinors}, and also used the trace to rewrite the result as an inner product (see \eqref{eq:tracescalar}). 
We have also made use of \eqref{1-formwedgePsi} to convert the derivative into an exterior derivative, and the fact that  upon acting with $\eta^2_\mp \eta^{1 \, \dagger}_-$ on the left and taking the trace, only one of the terms in \eqref{1-formwedgePsi} survives due to \eqref{onegamma}.

Putting everything together, the mass contribution from the derivative term gives 
\begin{equation}
m_\lambda^d= \mp \dfrac{i}{8 \pi} \int_\Sigma e^{-\phi}\text{ Re }\Psi_1   \,  \left( \Psi_2 \cdot  d\, \text{Re} \Psi_1  \right) \  .
\end{equation}

\paragraph{Dilaton gradient term:} By performing the dimensional reduction and using \eqref{onegamma} one finds that 
\begin{equation}
\bar{\theta} \ \underline{d \phi } \  \theta  =\frac{e^{-5A}}{16\pi^2} \left[    \bar{\lambda}_- \lambda_+\otimes \left(\eta^{1\dagger}_-  \underline{d \phi } \  \eta^1_+ +\eta^{2\dagger}_\pm  \underline{d \phi } \   \eta^2_\mp \right)\right] + \, \mathrm{c.c.} =0  \quad \Rightarrow \quad   m_\lambda^{\phi}=0 \ .
\end{equation}
Consequently, the dilaton gradient does not contribute to the 4d gaugino effective mass. The same argument holds for terms involving derivatives of the warp factor.

\paragraph{RR flux term:} At first sight, the operator in \eqref{eq:operators} involving  the RR fluxes splits them according to the amount of world-volume indices, denoted as $n$. We get 
 \begin{equation}\label{eq:usefulID}
\Gamma^\alpha \underline{F_q}\Gamma_\alpha- \frac{1}{2} \Gamma^M \underline{F_q} \Gamma_M =(-1)^q\sum_{n=0}^q (p+q-2n-4)\underline{F_q^{(n)}} \ .
\end{equation}
The coefficient in front of each component depends on a peculiar combination of the degree $q$, the number of world-volume indices $n$ and the dimension of the D$p$-brane involved. However, we will see that in the final expression this odd-looking separation of the different components will not appear. In order to show this, let us first focus on a particular component  of the purely internal fluxes $\hat{F}_q^{(n)}$. Its mass contribution is proportional to 
\begin{equation} \label{intermezzoF}
    (-1)^q \bar{\theta} \underline{\hat{F}_q^{(n)}} {\cal{P}}_q \theta = \frac{e^{-5A}}{8\pi^2} \overline{\lambda}_-\lambda_+ \otimes 
    \eta^{1 \, \dagger}_- \underline{\hat{F}_q^{(n)}} \eta^2_\mp
    + \mathrm{c.c.} 
    = \pm \frac{ie^{-4A}}{8\pi^2} (\lala)^* 
    \left(\Psi_2 \cdot \hat{F}_q^{(n)}\right) + \mathrm{c.c.}
\end{equation}
where, starting from the second equality, the contraction is performed using the internal $\gamma$-matrices, and in the last one we   used \eqref{eq:tracescalar}. The  last expression provides the key to understand why the apparent separation in components of equation \eqref{eq:usefulID} does not appear in the final expression: for a calibrated brane many components of $\Psi_2$ vanish on the brane locus, as we saw in \eqref{eq:seispurespinors}, and only those satisfying $p+q-2n=6$ give a contribution.

A similar story holds for the RR fluxes with external legs. Note that the relevant combination is now $ (\mathrm{vol}_4 \wedge  e^{4A} \tilde{F})_q^{(n)}$, so the degree of these forms is at least 4. Also, because we consider space-filling D$p$-branes   these fluxes have $n\geq 4$.  Performing the dimensional reduction this time we find 
 \beq\label{eq:abcd2}
 (-1)^q \bar{\theta}\,  \underline{(\mathrm{vol}_4 \wedge  e^{4A} \tilde{F} )_q^{(n)}}   \, \mathcal{P}_q   \theta = i (-1)^{1+q} \ \bar{\theta} (\gamma_{(5)}^{q+1} \otimes \underline{  \tilde{F}^{(n-4)}_{q-4}}   ) \theta =  i  \frac{e^{-5A}}{8\pi^2}  (\lambda \lambda)^* \otimes (   \eta^{1\dagger}_-    \tilde{F}^{(n-4)}_{q-4} \eta_\mp^2) + \mathrm{c.c.} 
  \eeq
so the main difference with the previous case comes from the appearance of the 4d chirality matrix.  As it happened for internal fluxes, $\tilde{F}$   also appears in a contraction with $\Psi_2$ and thus is subject to the projections in \eqref{eq:seispurespinors}. Nevertheless, for fluxes with 4d indices we find that the ones making a contribution are those with $p+q-2n=2$. 
As a consequence,  we see that all fluxes that make a contribution to the mass term have the same coefficient in \eqref{eq:usefulID} but there is a relative sign between internal fluxes and those with external indices. This means that the   combination of fluxes appearing on the effective mass is
\begin{equation}
{\hat F}-i\tilde{F}=  (1+i*_6\alpha)\hat{F} \ 
\end{equation}
after using the duality condition \eqref{sdF6}. Moreover, because fluxes appear in  the mass term contracted with the pure spinor $\Psi_2$, which satisfies the self-duality condition \eqref{*Psi}, we conclude that the contribution from the $\hat{F}$ and $\tilde{F}$ fluxes is exactly the same, so we will write the final result in terms of internal fluxes $\hat{F}$ only. 

Finally, we put  everything together to find that the total RR flux contribution to the gaugino mass term is given by 
 \begin{equation}
m^F_\lambda=\mp \dfrac{1}{8\pi}\int_\Sigma e^{-\phi}\text{ Re }\Psi_1  (\Psi_2 \cdot   e^{\phi} {\hat F} ) \ ,
 \end{equation}
where one must take into account that the calibration conditions select only a few components of this flux, that we list in Table \ref{table1}. 

\paragraph{NSNS flux term:} Using the same notation as above, we find 
\begin{equation} \label{Hcombination}
\Gamma^\alpha\underline{H_{\alpha}}-\underline{H}  =   2\underline{H}^{(3)}+\underline{H}^{(2)} -\underline{H}^{(0)} \ . 
\end{equation}
In the double-spinor notation, this is combined with the Pauli matrix $\sigma^3$, which gives
\beq \label{eq:intermediaH}
    \bar{\theta} \underline{H}^{(n)} \sigma_3 \theta = \frac{e^{-5A}}{16 \pi^2} \overline{\lambda}_- \lambda_+
    \otimes \left( 
    \eta^{1 \, \dagger}_- \underline{H}^{(n)} \eta^1_+ 
    - 
    \eta^{2 \, \dagger}_\pm \underline{H}^{(n)} \eta^2_\mp
    \right) + \mathrm{c.c.}  
\eeq 
Using once again the calibration condition \eqref{eq:kappaintern} we find that 
\begin{equation}
\eta^{2 \, \dagger}_\pm \underline{H}^{(n)} \eta^2_\mp=(-1)^{1+n } \ \eta^{1 \, \dagger}_- \underline{H}^{(n)} \eta^1_+
\end{equation}
so only the  components with zero and two  world-volume indices ($H^{(0)}$ and $H^{(2)}$) make a contribution to the effective mass. We can furthermore proceed analogously to the previous cases and re-write these contributions in terms of the pure spinors. We get
\beq\label{eq:Hterm}
    \bar{\theta} \underline{H}^{(n)} \sigma_3 \theta= \mp
    \frac{e^{-4A}}{8\pi^2} (\lala)^* \left[
    \Psi_2 \cdot \left( {H}^{(n)} \wedge \re \Psi_1 \right)
    \right] + \mathrm{c.c.}
\eeq
where we have used \eqref{HwedgePsi}, and the fact that only one of their terms survives in the relevant trace.

%
Combining everything together we get 
\begin{equation}
m_\lambda^H= \mp \dfrac{i}{8 \pi}
\int_{\Sigma_{p'}} e^{-\phi}\text{Re} \Psi_1  
\left[ \Psi_2 \cdot 
\left(({H}^{(2)}-H^{(0)}) \wedge \text{Re} \Psi_1 \right)\right]  
 \  .
\end{equation}   

An alternative computation of the H-flux can be done by using \eqref{eq:kappaintern} on the spinor on the left in \eqref{eq:intermediaH}    
\begin{equation}
  \gamma_{Dp}\left( \underline{H_3^{(0)}}-\underline{H_3^{(2)}}  \right) = \underline{ \star_\Sigma \left( H_3^{(0)}+ H_3^{(2)}  \right)} \ ,
   \  \label{eq:H-decomposition}
 \end{equation}
here $\star_\Sigma$ is the Hodge duality operator on the cycle $\Sigma_{p'}$ whose definition is in \eqref{eq:hodgedualcycle}. This implies that the H-flux contribution to the gaugino mass term can also be written as  
\begin{equation} \label{mHalternativo}
m_\lambda^H= \dfrac{-i}{8\pi}\int_{\Sigma_{p'}} e^{-\phi} \text{ Re }\Psi_1 \left[ \left( \star_\Sigma H_3\right) \cdot\Psi_{2} \right] \ .
\end{equation}  
As in previous cases, the calibration conditions automatically projects out  some of the flux contributions.

 The appearance of the relative sign between the two contributions of the NSNS flux may look surprising, specially since one does not find this feature in the RR flux sector. Let us briefly explain why this relative sign appears only on the NSNS flux mass. Recall that in the RR sector, we split  the flux according to \eqref{eq:usefulID}, and we showed that the calibration condition implies that only RR fluxes satisfying $p+q-2n=6$ mod $4$ lead to a mass. Only when demanding a four-dimensional Lorentz-invariant background we were forced to choose $p+q-2n=6$. If we relax this condition, RR fluxes satisfying $p+q-2n=2$   contribute to the mass terms as well.  This would result in another contribution to the mass from each flux, with two more worldvolume indices as compared to the ones we previously considered. Note also    that  \eqref{eq:usefulID} indicates that these new mass contributions have the opposite sign to the ones we were interested in.  For example, on a 10d flat background, the gaugini on a D3-brane can acquire a mass from $F_3^{(2)}$ with the opposite sign to the mass from $F_3^{(0)}$.  This property turns out to be completely general:   any   flux can lead to two contributions to the gaugino mass, with a sign difference between them. It is when  imposing extra constraints, such as compatibility with a given background,  that some of the contributions must  be ruled out. In our case, the background only allows for one of these contributions in the RR flux sector, while in the NSNS flux sector the background allows for mass contributions with either/both sign(s).

\subsection{Dimensional reduction: summary of results and analysis}
 
Putting together our results so far, we see that the four-dimensional action for the gaugino kinetic and mass terms takes the form 
\beq
{\cal S}_{\lambda\lambda}= \int d^4x \, \left( \frac{i}{2}f\,  \bar{\lambda}_+\gamma^\mu\nabla_\mu\lambda_+ + \frac{1}{2} m_{\lambda} \,  
\bar{\lambda}_-\lambda_+ + \mathrm{ c.c. } \right) \nonumber
\eeq
where the gauge kinetic function is given in \eqref{fkin} and the gaugino effective mass is
\begin{equation} \label{mlambda}
m_{\lambda}=\mp \, \dfrac{1}{8\pi}  \int_{\Sigma}  \text{Re } \Psi_1  \left[ \Psi_2 \cdot \left( \hat{F} + i d_{\Hb} (e^{-\phi} \re \Psi_1 )  \right)\right] 
\end{equation}
and
\beq \label{dHb}
d_{\Hb} \equiv d +(H^{(2)} -H^{(0)})  \wedge 
\eeq
Finally, we use the scalar delta-function $\delta^{(0)}$ defined in \eqref{delta0calibrated}, that is, the scalar version of the $(6-p')$-form representing the Poincar\'e dual of $\Sigma_{p'}$, and the Mukai pairing \eqref{Mukai}, to rewrite this effective mass in terms of an integral over the full six-dimensional manifold. In this way, we obtain the main result of this paper: 
\begin{equation}
m_{\lambda}= -\dfrac{i}{8\pi} \int_{M_6} \delta^{(0)}  \left\langle \Psi_{2} ,
 \hat{F}+ i d_{\Hb} (e^{-\phi} \re \Psi_1) 
 \right\rangle \ .
\label{mfinal}
\end{equation}

There are several features of this result that are worth pointing out. First, due to \eqref{eq:seispurespinors}, the (algebraic) calibration condition implies that several terms do not contribute to $m_\lambda$.  In Table \ref{table1} we put together all fluxes that can give a contribution to the effective mass. As for the derivative, the only non-zero contributions come from the derivative along world-volume directions, as the original formula \eqref{der1} indicates. 

Note that the components of RR fluxes that give a contribution to the gaugino mass are compatible with the $T$-duality rules. In fact, the components surviving the projection can be guessed by starting from one of the cases at the extrema. It turns out that the extremum on the low part of the Table \ref{table1} is known. Indeed, in a compactification with D9-branes, one must necessarily introduce O9-planes, so this is a Type I compactification. Therefore, the only flux available is the RR 3-form $F_3$. This is precisely what we find from the calibration condition. Obviously, the only possibility is that all indices of $F_3$ lie along the world-volume of the D9-brane, namely $n=3$. A T-duality turns the D9 into a D8-brane, and $F_3$ with $n=3$ into both $F_2$ with $n=2$ and $F_4$ with $n=3$, in agreement with what we find. The rest of the table can be obtained by iterating this argument.

\begin{table}
\begin{center}
 \begin{tabular}{|c||c|c|c|c|c|c|c||c|} 
\hline
Brane $ \diagdown$ Flux & ${F}_0$ & ${F}_1$ & ${F}_2$ & ${F}_3$ & ${F}_4$ & ${F}_5$ & ${F}_6$     & $H $  \\ \hline\hline
D3  &  & - &  & n=0 &  & - &    & n=0\\ \hline
D4  & - &  & n=0 &  & n=1 &  & -   &n=0 \\ \hline
D5  &  & n=0 &  & n=1 &  & n=2 &   & n=0 $\&$ 2\\ \hline 
D6  & n=0 &  & n=1 &  & n=2 &  & n=3  & n=0 $\&$ 2\\ \hline
D7  &  & n=1 &  & n=2 &  & n=3 &   & n=2\\ \hline
D8  & - &  & n=2 &  & n=3 &  & - &  n=2\\ \hline
D9  &  & - &  & n=3 &  & - &  &  - \\ 
\hline 
\end{tabular}
 \end{center} \caption{Flux that originate potentially non-vanishing contributions to the gaugino mass. The dashes indicate components that are projected out due to the calibration of the cycle.}  \label{table1}
\end{table}


Furthermore, we note that in \eqref{mfinal} both the RR fluxes and the derivative term contribute to the effective mass of the gaugino exactly in the same way as they appear in the superpotential \eqref{W10D}  \cite{Grana:2006hr,Benmachiche:2006df}. This was conjectured in \cite{Kachru:2019dvo} for the simpler situation of D7-branes in an SU(3) structure background. However, the full mass term is not proportional to the (integrand of) the superpotential: somewhat counterintuitively, the contributions generated by $H^{(0)}$ and $H^{(2)}$ enter with opposite sign. More precisely, the component of  $H$ flux with all indices transversal to the D-brane world volume appears with the \textit{opposite} sign as compared to \eqref{W10D}. This relative sign constitutes a crucial consistency check. Indeed, for the simple case of calibrated D3-branes (where the structure at the location of the brane must be SU(3), corresponding to the spinors in \eqref{SU3ps}) it is well-known that the gaugino mass is proportional to $\left(\bar{G}_3\cdot \Omega \right)$ \cite{Grana:2002tu,Camara:2003ku,Grana:2003ek}. In contrast, and as will be described in some detail below,  for D7-branes in a similar configuration we know that the gaugino mass is given by $\left(G_3\cdot \Omega \right)$ \cite{Camara:2004jj,Jockers:2004yj}. Consistency with these two particular examples would be hard to achieve without this odd-looking relative sign in the definition of $\tilde{H} = H^{(2)}-H^{(0)}$. As a consequence, the combination $\left(G\cdot \Psi_2 \right)$ appearing in the superpotential gives the gaugino mass  only for D7-, D8- and D9-branes, for which $H^{(0)} = 0$. This  relative sign is  absorbed in the alternative expression for the H-flux contribution to the gaugino mass involving the Hodge star on the cycle, Eq. \eqref{mHalternativo}. Writing this in terms of an integral over the whole manifold, and combining with the other contributions, this is
\begin{equation}
m_{\lambda}= -\dfrac{i}{8\pi} \int_{M_6} \delta^{(0)}  \left\langle \Psi_{2} \ , \
 \hat{F} +  i e^{-\phi} (\pm \star_{\Sigma} H + d  \, \re \Psi_1) 
 \right\rangle \ .
\label{mfinalalternativo}
\end{equation}

\medskip

\paragraph{Supersymmetric backgrounds:}
Finally, let us briefly show that the gaugino mass terms we have computed vanish for calibrated D-branes in supersymmetric backgrounds, as they should. For this it is more convenient to go back to the original brane action in terms of the internal spinors $\eta_{1,2}$ as in \eqref{SF}. Recall that the mass terms arise from the operator \eqref{eq:operators}, that is a combination of the supersymmetry variation of the gravitino  \eqref{eq:operator1} and the dilatino \eqref{eq:operator2}.  In supersymmetric backgrounds these variations vanish, and upon using the warped compactification ansatz \eqref{epsilon12}, one finds  that internal spinors must satisfy the following conditions: 
\begin{eqnarray} 
\label{eq:operator1b}
\left(\nabla_m +\dfrac{1}{4} \underline{H_{m}} \right) \eta^1_+
\mp\dfrac{e^{\phi}}{16}  \sum_q \underline{F_q} \gamma_m   \eta^2_{\mp} &=& 0 \ , \\
\left(\underline{d\phi} +\dfrac{1}{2} \underline{H} 
\right) \eta^1_+ - \dfrac{e^{\phi}}{16}  \sum_q  \underline{F_q} (10-2q)  \eta^2_\mp &=& 0 \ .  \label{eq:operator2b}
\end{eqnarray} 
Note that on both expressions we only look at the  internal component and thus all contractions are taken with internal $\gamma$ matrices.  Analogous conditions hold upon exchanging the spinors and shifting the signs for the corresponding fluxes accordingly. 

In order to proceed, recall   the minus sign in  the definition of $\theta_2$ as compared to the supersymmetry parameter $\epsilon_2$.\footnote{There is also a warp-factor difference between both definitions but this does not modify our procedure. The reason is that warp factor gradients do not contribute to the gaugino mass, same as the dilaton gradients.} This relative sign implies that the above SUSY conditions can be used to re-write the multiple  contributions to the gaugino mass in terms of
the RR flux contribution, and as seen above it is sufficient to focus on the internal terms, characterized by $\hat{F}$. 
Three types of mass contributions exist: the ones coming from the four-dimensional part of $D_\mu$,  twice (after imposing the supersymmetry condition \eqref{eq:operator1b}) the contribution from the cycle directions $D_\alpha$, and twice (after imposing \eqref{eq:operator2b}) the contribution from $\Delta$ with the relative (-1/2) factor from the combination in \eqref{eq:operators}. In order to sum all contributions it is convenient to use identities arising from the anticommutation of $\Gamma$ matrices similar to \eqref{eq:usefulID}. We find that the sum of all contributions is proportional to
\begin{equation}
m_\lambda^{\rm SUSY}\propto    \sum_q    (-1)^{\left[\frac{q-1}{2}\right]} \left[ 4 + 2 (p-3-2n)-(10-2q)\right]\, \eta^{1\dagger}_-  \underline{\hat{F}_q^{(n)}} \eta^2_{\mp} \ .
\end{equation}
This combination vanishes due to  \eqref{eq:seispurespinors}. As expected, no gaugino mass terms are induced for calibrated branes when the background is supersymmetric. 

\subsection{D-branes in SU(3) structure}
\label{sec:DbranesSU3}

Here we compute the gaugino masses for D$p$ branes in an SU(3) structure-background. First, note that the calibration condition \eqref{eq:kappaintern} implies that one cannot have D4- or D8-branes in this context. In contrast, D9-branes can only be present for an internal manifold with SU(3) structure, while D3-branes also require an SU(3) structure but only at the location of the brane\footnote{This means that D3-branes can be calibrated in a so-called dynamic SU(2) structure, as long as the structure reduces to SU(3) (the internal spinors $\eta^1$ and $\eta^2$ become parallel) at the location of the brane.}.
However, since the computation of the gaugino mass is performed at the location of the brane, the SU(3) structure expression gives the general gaugino mass term for D3-branes. 

The SU(3) structure pure spinors are given in  \eqref{SU3ps}\footnote{The gaugino masses should actually have an overall extra phase $e^{-i \theta}$ from $\Psi_2$, that we are not writing. This extra phase is important in AdS$_4$ compactifications since it should be aligned with the phase in $\mu$ \cite{Bena:2019mte}.}.  We first study the type IIB branes.

\begin{itemize}
    \item For a D3-brane ($\theta=0$), the gaugino mass \eqref{mlambda} reduces to 
\beq 
m_{\lambda}^{D3}= \dfrac{1}{8\pi} \int_{M_6} \delta^{(0)}     ({F_3}- ie^{-\phi} H ) \wedge \Omega  = \dfrac{1}{8\pi} \int_{M_6} \delta^{(0)}      \bar G_3  \wedge \Omega  \ , 
\eeq
where the contribution of the NSNS flux is from $H^{(0)}$ only, and in the last equality we have used the usual definition $G_3 = F_3 + i e^{-\phi} H$. This is the exactly the mass-term computed in  \cite{Grana:2002tu} for the particular case of Calabi-Yau manifolds. Note that the exterior derivative gives no contribution, and this extends to the more general case of SU(3)$\times$SU(3) structure, since the latter has to be an SU(3) structure at the location of the D3-brane.  

\item For D5-branes ($\theta=-\pi/2$), we get   
\beq \label{D5SU3}
m_{\lambda}^{D5}= \dfrac{-i}{8\pi} \int_{M_6} \delta^{(0)}   ( {F_3}+ i e^{-\phi} dJ) \wedge \Omega   \ .
\eeq
The other RR fluxes do not give a contribution as $\Psi_2$ has only a 3-form piece. We can see here clearly that the projection onto the component with one world-volume index as indicated in Table \ref{table1} is redundant as only the component of $F_3$ with three anti-holomorphic indices enters, out of which only one can be world-volume index for a calibrated D5-brane for which $\Sigma$ should be a holomorphic cycle.  Finally, the NSNS fluxes give no contribution as $H\wedge \re \Psi_1$ is a five-form. In this case the integrand is proportional to that of the superpotential.

\item For D7-branes ($\theta=0$), we get  
\beq
m_{\lambda}^{D7}= \dfrac{1}{8\pi} \int_{M_6} \delta^{(0)}     (   { F_3}+ ie^{-\phi} H ) \wedge \Omega 
= \dfrac{1}{8 \pi} \int_{M_6} \delta^{(0)}      G_3  \wedge \Omega  
\eeq
as in Calabi-Yau manifolds \cite{Camara:2004jj,Jockers:2004yj}. The derivative of the (real part of) the pure spinor $e^{-\phi} e^{iJ}$ gives no contribution as it has one and five-form pieces only. 
The RR part is straightforward. The $H$ contribution comes from the $H^{(2)}$ piece, and thus has the opposite sign as for D3-branes. Once again, the components of $F_3$ and $H$ that give a non-zero contribution can only have two world-volume indices. 

\item For D9-branes ($\theta=-\pi/2$), we get the same result as for a D5-brane \eqref{D5SU3}, except that the contribution is not localized. 

\end{itemize}

\noindent Finally, for type IIA the only brane one can have in SU(3) structure is a D6-brane. 

\begin{itemize}
    \item For D6-branes 
    we can re-absorb the phase in $\Omega$ and thus simply take $\Psi_1 = i\Omega$.       Thus, we obtain 
\beq
m_{\lambda}^{D6}= \dfrac{i }{8 \pi} \int_{M_6} \delta^{(0)}    \left[ \<  { F}, e^{iJ} \> + d (e^{-\phi} \re \Omega) \wedge J + i e^{-\phi} (H^{(2)}-H^{(0)}) \wedge \re \Omega \right]   \ , 
\eeq
Here all NSNS terms (derivative, $H^{(2)}$ and $H^{(0)}$) contribute, and we cannot combine them with the RR piece to form either a $G$ or a $\bar G$ flux. 

\end{itemize}

\acknowledgments
We thank Severin L\"ust, Jamie Rogers, Radu Tatar, Flavio Tonioni, Vincent van Hemelryck, Thomas van Riet and  specially Luca Martucci for interesting comments and valuable insights.
This work was partially supported
by the ERC Consolidator Grant 772408-Stringlandscape, the ANR grant Black-dS-String. The work of N.K. is supported by the Leverhulme Trust under grant no RPG-2018-153, and additionally by the National Agency for the Promotion of Science and Technology of Argentina (ANPCyT-FONCyT) Grants PICT-2017-1647 and PICT-2015-1525.

\newpage 

\appendix
\section{Conventions, definitions and useful identities}\label{app:coventions}

\subsection{Spinor conventions}
\label{sec:spinorconventions}

We use conventions where the standard intertwiners $A, C$ and $D$ defined as ($t$ is the amount of temporal directions) 
\begin{equation} \label{intertwinners}
  (-1)^{t} \Gamma^\dag_M=A\Gamma_M A^{-1} , \qquad
   -\Gamma_M^T=C^{-1}\Gamma_M C , \qquad   (-1)^{t-1}\Gamma^*_M=D^{-1}\Gamma_M D \ .
\end{equation} 
take the form 
\begin{equation}
    D=1\ , \ A=C=\Gamma^0.
\end{equation} 
Thus, the different conjugates are written as 
\begin{equation}
   \bar{\theta}\equiv\theta^\dagger A=\theta^\dagger \Gamma^0 , \qquad
   \theta^t \equiv \theta^TC^{-1}=-\theta^T \Gamma^0 \ , \quad \theta^c\equiv D \theta^*=\theta^* \ , 
\end{equation}
where no double spinor notation is intended. 
In particular, Majorana spinors such as $\theta^1$ and $\theta^2$ satisfy $\theta^c=\theta^*=\theta$.

The decomposition \eqref{gamma10} implies that the 4d intertwiners are analogous to the ten-dimensional ones, i.e.  $D^{(4)}=1$, $C^{(4)}=A^{(4)}=\gamma^0$. As for the (euclidean) 6d spinors, our decomposition implies that the intertwiners must be $C^{(6)}= A^{(6)} = D^{(6)} = 1$. These imply $\gamma^m$ are imaginary anti-symmetric. 

A representation of 4d matrices compatible with our conventions has 
\begin{equation}
\gamma^0=\left(\begin{matrix}
0 & \sigma^3 \\ -\sigma^3 & 0
\end{matrix} \right) \quad , \quad \gamma_{(5)}=\left(\begin{matrix}
0 & -i\sigma^1 \\ i \sigma^1 & 0
\end{matrix} \right) \ .
\end{equation}
In this representation a four-dimensional Dirac spinor with positive chirality must be of the form 
\begin{equation}
\lambda_+=\dfrac{1}{\sqrt{2}}\left(\begin{matrix}
 -i \sigma^1\bar \psi \\ \bar \psi 
\end{matrix} \right)
\end{equation}
with $\bar \psi$ a two-component complex  spinor. This implies that the bilinear appearing in the usual mass term is 
\begin{equation} \label{ll}
(\lambda \lambda)^* \equiv \bar{\lambda}_-\lambda_+  =i\bar \psi_{\dot \alpha} \bar \psi_{\dot \beta} \epsilon^{\dot \alpha\dot \beta} 
\end{equation}
Finally, let us provide some useful identities.  For any  
pair of chiral fermions $\eta^i_\pm$ in 6d we have
\begin{eqnarray}
\label{eq:Majoranazero}
\eta_\pm^{i\dagger}\gamma^{m_1 ... m_n} \eta_\mp^j &=& 0\quad \quad i,j=1,2 \quad, \quad  n \, \mathrm{even} \ , 
\\ 
 \eta_\pm^{i\dagger}\gamma^{m_1 ... m_n} \eta_\pm^j &=& 0\quad \quad i,j=1,2 \quad, \quad  n \, \mathrm{odd} \ , \\
  \eta_\pm^{i\dagger}\gamma^{m} \eta_\mp^i &=& 0 \ . \label{onegamma}
\end{eqnarray}

\subsection{Other conventions and definitions }

Throughout the paper, an underlined differential form denotes the contraction of said form with the appropriate (antisymmetrized) product of gamma matrices. Note that depending on the context these can be 10d or 6d matrices. Writing the matrices generically as $\Gamma^M$, the precise definition is  
\beq\label{eq:def-underline}
\underline{F_q}\equiv\frac{1}{q!} F_{M_1...M_q} \Gamma^{M_1...M_q} \ .
\eeq
An underlined form with one explicit index implies that the given index is not contracted with a gamma matrix, for instance
\beq \label{HM}
 \underline{H_{m}}\equiv \frac12 H_{m n_1 n_2} \gamma^{n_1 n_2} \ .
 \eeq
Given differential form on $M_6$, the different components can be computed as  
\begin{equation}
    A_{m_1 \dots m_q} = \frac{1}{8} \Tr \left[
    \underline{A} \gamma_{m_q \dots m_1}
    \right]. 
    \label{traza}
\end{equation}
The scalar product $\cdot$ contracts $p$-forms as follows: 
\beq \label{eq:scalarproduct}
A_p\cdot B_p \equiv \dfrac{1}{p!}A_{m_1...m_p}B^{m_1...m_p}.
\eeq
The generalization for polyforms is simply the sum of scalar products involving the different components.  A useful identity involving the scalar product of (poly)forms is
\begin{equation}\label{eq:scalar.wedge}
\left(A_p\cdot B_p\right) \mathrm{vol}_6 = A_p\wedge *_6 B_p \ , 
\end{equation}
where $\mathrm{vol}_6$ is the volume form on $M_6$. Thus, given a polyform $\Psi$ satisfying the self-duality condition \eqref{*Psi} and another generic one, say $\Phi$, this means that 
\begin{equation}
    \langle \Psi,\Phi\rangle = \mp i \left(\Psi \cdot \Phi\right) {\rm vol}_6
\end{equation}
after using that $\alpha \, *_6 = \mp *_6 \alpha$.

\noindent The operators in \eqref{SF} that involve the world-volume flux are given by 
\begin{eqnarray}
\label{eq:LF}
L_{Dp} \left(\mathcal{F}\right) &=& -\Gamma_{\kappa}\sum_{r\geq 1} \left(\pm \sigma_3\right)^r\left( \mathcal{F}^r\right)^{\alpha\beta}\Gamma_\alpha D_\beta \ , \\ \Gamma_\kappa \left(\mathcal{F}\right)&=&\left(\begin{matrix}
0&\Gamma_{Dp}^{-1}\left(\mathcal{F}\right)\\ \Gamma_{Dp}\left(\mathcal{F}\right)& 0 
\end{matrix}\right) \ , \label{eq:LF2}
\end{eqnarray}
where $\Gamma_\kappa$ is the relevant operator for $\kappa$-gauge fixing in the spinor doublet notation, and satisfies
\begin{equation}
    \Gamma_{Dp}^{-1}=\Gamma_{Dp}^\dagger=(-1)^{1+p+\left[\frac{p}{2}\right]}\Gamma_{Dp}.
\end{equation}

\noindent In order to define the Hodge star operation on the cycle $\Sigma$ wrapped by the brane used in Section \ref{sec:WVaction}, we first need to impose that indices in any form are ordered as follows   to remove sign ambiguities:
\begin{equation}
A_q^{(n)}=\dfrac{A_{a_1 ...a_n m_1...m_{q-n}}}{n!(q-n)!} \ dy^{a_1}\wedge... \wedge dy^{a_n}\wedge dy^{m_1}\wedge... \wedge dy^{m_{q-n}}  
\end{equation}
and the Hodge star operator is then ($\varepsilon_{1...p'}=1$)
\begin{equation}\label{eq:hodgedualcycle}
\star_\Sigma A_q^{(n)}= \dfrac{\sqrt{g} \ \varepsilon_{a_1 ... a_{p'}} A^{a_{p'-n+1} ... a_{p'}}{}_{m_1...m_{q-n}} }{n!(q-n)!(p'-n)!} \ dy^{a_{1}}\wedge...\wedge dy^{a_{p'-n}}\wedge dy^{m_1}\wedge ... \wedge dy^{m_{q-n}}\ .
\end{equation}

\subsection{Useful identities}

We provide an example of a very useful type of identities to perform the dimensional reduction  \begin{equation}\label{eq:tracescalar}
 \eta_\pm^{2\dagger}\underline{F_q} \eta_+^1=\text{Tr} ( \underline{F_q} (\eta_+^1\eta_\pm^{2\dagger})) =\dfrac{i|\eta|^2}{8} \text{Tr}(\underline{F_q }\ \underline{\Psi_{q,\pm}}) = i|\eta|^2 (-1)^{\frac{q(q-1)}{2}}  F_q \cdot\Psi_{q,\pm} \ .
\end{equation}
In the second equality we used the definition of pure spinors \eqref{purespinors}, and $\Psi_{q,\pm}$ denotes the $q$-form component of $\Psi_{\pm}$.

Using the $\Gamma$-matrix anticommutator  one can easily show that
\begin{equation}\label{eq:usefulID2}
\Gamma^\alpha\underline{F_q}\Gamma_\alpha=(-1)^q\sum_{n=0}^q(p+1-2n)\underline{F_q^{(n)}} \quad  , \quad  \Gamma_{Dp}\underline{F_q^{(n)}}=(-1)^n \underline{F_q^{(n)}} \Gamma_{Dp}
\end{equation}
 
The bispinor operation that corresponds to the wedge of a 1-form (e.g. $F_1$) with the pure spinor $\Psi_1$ is given by
\begin{equation} \label{1-formwedgePsi}
    \underline{F_1 \wedge \Psi_1} = -4i e^{-A} F_m \left[\gamma^m, \left(\eta^1_+ \eta^{2 \, \dagger}_\mp
    \right)\right]_\mp \ ,
\end{equation}
where $[,]_-$ ($[,]_+$) stands for the commutator (anti-commutator). 

The wedge of a 3-form (e.g. $H$) gives
\begin{equation} \label{HwedgePsi}
    \underline{{H}\wedge \Psi_1} = \frac{-i e^{-A}}{6!} {H}_{mnp}
    \left[ 
    \gamma^{mnp} \eta^1_+ \eta^{2\,\dagger}_\mp \mp
    3 \gamma^{mn} \eta^1_+ \eta^{2\,\dagger}_\mp \gamma^p + 
    3 \gamma^{m} \eta^1_+ \eta^{2\,\dagger}_\mp \gamma^{np} \mp
     \eta^1_+ \eta^{2\,\dagger}_\mp \gamma^{mnp}
    \right]
\end{equation}
 
\bibliographystyle{JHEP}
\bibliography{refs}

\end{document}